\documentclass[12pt]{iopart}
\usepackage{graphicx,amsfonts,amssymb,enumerate}
\usepackage[colorlinks, citecolor=blue, linkcolor=blue,
urlcolor=blue]{hyperref}

\newcommand{\beq}{\begin{equation}}
\newcommand{\eeq}{\end{equation}}
\newcommand{\beqa}{\begin{eqnarray}}
\newcommand{\eeqa}{\end{eqnarray}}
\newcommand{\ket} [1] {\vert #1 \rangle}
\newcommand{\bra} [1] {\langle #1 \vert}
\newcommand{\braket}[2]{\langle #1 | #2 \rangle}

\newcommand{\widebar}[1]{\overline{#1}}

\def\bra#1{\langle#1\vert}
\def\ket#1{\vert#1\rangle}

\def\Longarrow{\protect\@lra}
\def\@lra{\relbar\joinrel\relbar\joinrel\relbar\joinrel
          \relbar\joinrel\rightarrow}

\begin{document}

\title{Topological order on the Bloch sphere}

\author{Rotem Liss$^1$, Tal Mor$^1$ and Rom\'an Or\'us$^{2,3,4,5,6}$}

\address{$^1$ Computer Science Department, Technion, Haifa, 3200003, Israel}
\address{$^2$ Institute of Physics, Johannes Gutenberg University, 55099 Mainz, Germany}
\address{$^3$ Donostia International Physics Center, Paseo Manuel de Lardizabal 4, E-20018 San Sebasti\'an, Spain}
\address{$^4$ Ikerbasque Foundation for Science, Maria Diaz de Haro 3, E-48013 Bilbao, Spain}
\address{$^5$ Multiverse Computing, Pio Baroja 37, 20008 San Sebasti\'an, Spain}
\address{$^6$ Author to whom any correspondence should be addressed.}
\ead{roman.orus@dipc.org}

\begin{abstract}
A Bloch sphere is the geometrical representation of an arbitrary
two-dimensional Hilbert space. Possible classes of entanglement and
separability for the pure and mixed states on the Bloch sphere were
suggested by [M. Boyer, R. Liss, T. Mor, PRA {\bf 95}, 032308 (2017)].
Here we construct a Bloch sphere for the Hilbert space spanned by one of the ground states of Kitaev's toric code model and one of its closest product states. We prove that this sphere contains only one separable state, thus belonging to the fourth class suggested by the said paper. We furthermore study the topological order of the pure states on its surface and conclude that, according to conventional definitions, only one state (the toric code ground state) seems to present non-trivial topological order. We conjecture that most of the states on this Bloch sphere are neither ``trivial'' states (namely, they cannot be generated from a product state using a trivial circuit) nor topologically ordered. In addition, we show that the whole setting can be understood in terms of Grover rotations with gauge symmetry, akin to the quantum search algorithm.
\end{abstract}


\maketitle

\section{Introduction}
Topological order \cite{wen} (TO) is a type of order in quantum matter deeply rooted in quantum entanglement. As such, TO cannot be detected by local order parameters. Some typical signatures of TO are anyonic excitations, ground-state topological degeneracy, and indistinguishability of local reduced density matrices. Moreover, topologically-ordered systems form the basis for lattice gauge theories \cite{lgt} and have been proposed in several aspects of quantum computation, including the construction of topologically-protected qubits and topological quantum computation by anyonic braiding \cite{kita}. 

The study of robustness in TO has been considered from many angles. It is known that two-dimensional (2d) TO survives local Hamiltonian perturbations at zero temperature \cite{robustTO}, which leads to the notion of topological phases of matter. However, 2d TO does not survive at finite temperature \cite{tempTO}, at least formally. In practice, though, there is a trade-off between different physical scales (temperature, size of the system, correlation length), in such a way that TO may be kept approximately at finite temperatures under some circumstances. The question whether 2d TO survives many-body localization or not has been recently addressed, too \cite{TOMBL}. 

In this paper we show that 2d TO does not seem to survive under certain perturbations that can be characterized by a Bloch sphere. More specifically, we study a perturbation of one of the ground states of Kitaev's toric code \cite{kita} by constructing a Bloch sphere with this state and one of its closest product states \cite{closest}. Following a recent characterization of Bloch spheres \cite{bloch}, we prove that, independently of the chosen closest product state, the spheres constructed this way include only one product state, and probably include only one topological state. More precisely, our analysis indicates that the topological entanglement entropy changes continuously, and hints to the fact that the conventional definitions of topological order are not satisfied for such states. We also conjecture that those non-topological states are not ``trivial'' states in the sense defined by~\cite{BHV06,Hastings13}. We subsequently show in \ref{appendix} that one can conveniently write the whole setting in terms of Grover rotations with gauge symmetry, akin to the quantum search algorithm \cite{grover}.

\section{Setting}
We consider the ground states of Kitaev's toric code model \cite{kita}. This is a model of spins-1/2 (qubits) on the links of a square lattice, with Hamiltonian 
\beq
H_{TC}= - \sum_s A_s - \sum_p B_p , 
\eeq
with star $A_s$ and plaquette $B_p$ operators respectively defined as 
\beq
A_s \equiv \prod_{j \in s}\sigma_x^{[j]} ~~~~~~ B_p \equiv \prod_{j \in p}\sigma_z^{[j]}. 
\eeq
In the above equations, $s$ is a star and {$p$ is a plaquette on the two-dimensional square lattice, and $\sigma_\alpha^{[j]}$ is the Pauli matrix $\alpha$ at site $j$. The properties of this model are well-known. In particular, its ground state manifold has topological degeneracy: if the model is defined on a Riemann surface of genus $\mathfrak{g}$, then the ground state is $2^{2\mathfrak{g}}$-fold degenerate. The ground state subspace is also a stabilized space of $G$, the group of all the possible products of independent star operators. This group has size $|{G}| = 2^{n_s-1}$, where $n_s$ is the number of stars in the lattice. The $-1$ in the exponent is a consequence of the global constraint $\prod_s A_s = \prod_p B_p = \mathbb{I}$. Following \cite{hamma}, for a system with $n$ spins, one of the ground states of the model can always be written as 
\beq
\ket{\Psi_0} = \frac{1}{\sqrt{|{G}|}} \sum_{g \in {G}} g \ket{0}^{\otimes n}, 
\label{tcstate} 
\eeq
no matter what the topology of the underlying Riemann surface is.

As proven in \cite{closest}, the closest product states to $\ket{\Psi_0}$ are those of the form $g \ket{0}^{\otimes n}$, with $g \in {G}$. For simplicity we choose here the state $\ket{ \bar{0} }\equiv \ket{0}^{\otimes n}$ \footnote{Equivalent conclusions can be reached by choosing any of the other closest product states.}. This state, together with $\ket{\Psi_0}$, spans a 2-dimensional Hilbert space defining a Bloch sphere. An arbitrary pure   state in this space can be written as 
\beq
\ket{\theta, \phi} \equiv \cos{\left(\frac{\theta}{2}\right)} \ket{\bar{0}} + e^{i \phi} \sin{\left(\frac{\theta}{2}\right)} \ket{\bar{1}}, 
\label{esta} 
\eeq
where $\theta \in [0, \pi]$ and $\phi \in [0, 2\pi]$, and the state $\ket{\bar{1}}$ is defined as 
\beq
\ket{\bar{1}} \equiv \frac{1}{\sqrt{|{G}| - 1}}\sum_{g \in {G}, g \neq \mathbb{I}} g \ket{0}^{\otimes n}. 
\label{esta2} 
\eeq
This characterizes the surface of a Bloch sphere, as shown in figure \ref{fig1}, where $\ket{\bar{0}}$ and $\ket{\bar{1}}$ sit at opposite poles. The toric code ground state is then given by 
\beq
\ket{{\Psi_0}} =    \sqrt{\frac{1}{|{G}|}} \ket{\bar{0}} + \sqrt{1-\frac{1}{|{G}|}} \ket{\bar{1}} .
\label{gr1}
\eeq
Similarly, the state at the antipode of $\ket{\Psi_0}$ is given by 
\beq
\ket{\widebar{\Psi_0}} =  \sqrt{1-\frac{1}{|{G}|}} \ket{\bar{0}} - \sqrt{\frac{1}{|{G}|}} \ket{\bar{1}}.
\label{gr2} 
\eeq
The Bloch sphere is shown in figure \ref{fig1}. 

\begin{figure}
	\centering
	\includegraphics[width=0.6\linewidth]{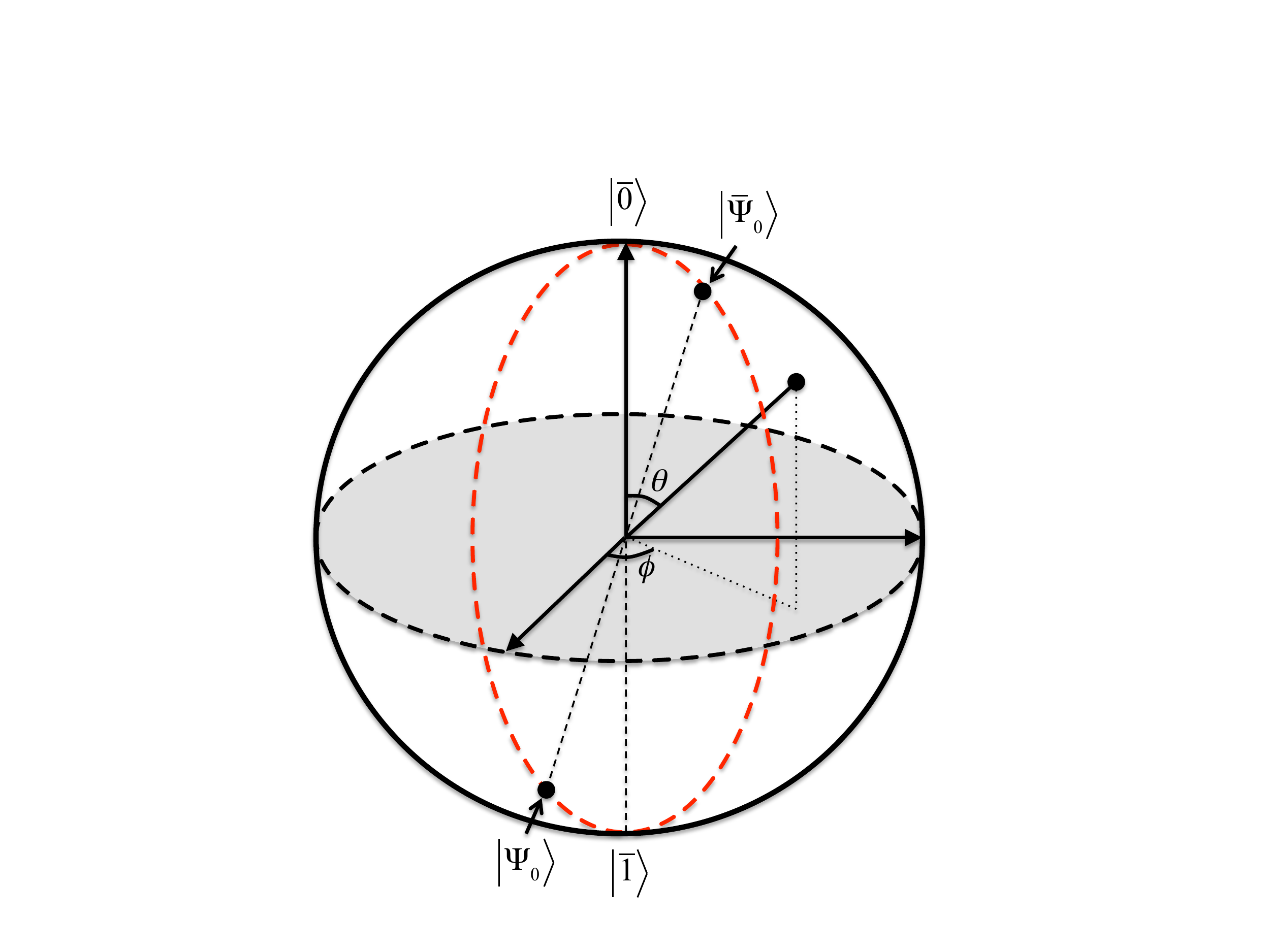}
	\caption{Bloch sphere defined by the toric code ground state $\ket{\Psi_0}$ and one of its closest product states $\ket{\bar{0}} = \ket{0}^{\otimes n}$, as specified in the text. The states $\ket{\widebar{\Psi_0}}$ and $\ket{\bar{1}}$ are the respective antipodal states. The different states on the surface of the sphere are characterized by angles $\theta$ and $\phi$. The interior corresponds to mixed states formed by convex combinations of the pure states on the surface.}
	\label{fig1}
\end{figure}

Let us now mention some general properties of the states built in this way. First, notice that the toric code ground state $\ket{\Psi_0}$ is actually obtained for angles $\theta_0 = 2 \arccos{(1/\sqrt{|{G}|})}$ and $\phi_0 = 0$. Therefore, we have a topologically-ordered pure state at this precise value of the angles. Second, equations (\ref{gr1}) and (\ref{gr2}) are identical to those of a Grover rotation in the quantum search algorithm \cite{grover}. We will come back to this property later. 

Two questions are now in order. First, which type of Bloch sphere are we dealing with, in terms of the classification in \cite{bloch}? And second, which states on the sphere have topological order? We can answer both questions by studying the entanglement of the pure states on the surface of the sphere. 

\section{Entanglement on the Bloch sphere}
To compute the entanglement properties, we first evaluate the reduced density matrix $\rho_A{(\theta,\phi)}$ of a subset $A$ of the spins defining the toric code Hamiltonian, for arbitrary values of $\theta$ and $\phi$. Following \cite{hamma}, let us call $B$ those spins that are not in $A$, and let us define the groups $G_{A,B}$ as the subgroups of $G$ consisting of all the products of star operators acting non-trivially on $A,B$, i.e., 
\beqa
G_A &\equiv& \left\{g \in G ~|~ g = g_A \otimes \mathbb{I}_B \right\}, \nonumber \\
G_B &\equiv& \left\{g \in G ~|~ g = \mathbb{I}_A \otimes g_B \right\}, \label{gAB}
\eeqa
with sizes $d_{A,B} = |G_{A,B}|$ respectively. Taking the partial trace over the spins in $B$ and using equations (\ref{tcstate})-(\ref{gr1}), we get 
\beqa
\rho_A{(\theta,\phi)} &\equiv& {\rm tr}_B \left( \ket{\theta, \phi} \bra{\theta, \phi} \right) = |a|^2 \ket{0_A}\bra{0_A} + |b|^2 \rho_A^{(0)} \nonumber \\ 
&+& \frac{1}{\sqrt{|{G}|}}\left(a^*b \ket{\phi_A} \bra{0_A} +  b^*a \ket{0_A} \bra{\phi_A}\right), 
\label{rhoa}
\eeqa
where $\ket{0_A} \equiv \ket{0}^{\otimes n_A}$, $n_A$ is the number of spins in $A$, $\rho_A^{(0)} = {\rm tr}_B\left( \ket{\Psi_0} \bra{\Psi_0} \right)$ is the reduced density matrix for the toric code ground state, and $\ket{\phi_A}= \sum_{g \in {G}_A} g_A \ket{0_A}$ is a non-normalized state~\footnote{We note that, according to equation~(\ref{gAB}), all operators $g \in G_A$ are of the form $g = g_A \otimes \mathbb{I}_B$.} The coefficients $a$ and $b$ are defined as 
\beqa 
a &\equiv&  \cos{\left( \frac{\theta}{2} \right)}- \frac{e^{i \phi}\sin{\left( \frac{\theta}{2} \right)}}{\sqrt{|G|-1}}, \nonumber \\
b &\equiv& e^{i\phi}\sin{\left( \frac{\theta}{2} \right)} \sqrt{\frac{|G|}{|G|-1}}. 
\label{ab} 
\eeqa
Moreover, we know from theorem 1 in \cite{hamma} that 
\beq
\rho_A^{(0)} = \frac{1}{f} \sum_{g \in G/G_B, \tilde{g} \in G_A} g_A \ket{0_A} \bra{0_A} g_A \tilde{g}_A = \frac{d_A}{f} P_A, 
\eeq
where $f = |G|/d_B$, and $P_A$ is a projector on a subspace of dimension
$f d_A^{-1}$ (namely, $P_A$ is a linear operator which satisfies
$P_A^2 = P_A$, and its image is a vector subspace
of dimension $f d_A^{-1}$~\footnote{This result comes from the proof
of theorem 1 in \cite{hamma}, which shows
$\left(\rho_A^{(0)}\right)^2 = f^{-1} d_A \rho_A^{(0)}$; thus, we
deduce $P_A^2 = P_A$ for $P_A \triangleq (f d_A^{-1}) \rho_A^{(0)}$,
and due to normalization, we deduce $\tr(P_A) = f d_A^{-1}$.
Therefore, $P_A$ is a projector to some vector space of dimension
$f d_A^{-1}$.}).

From our expression for $\rho_A{(\theta,\phi)}$ in equation~(\ref{rhoa}) we can now compute the 2-R\'enyi entropy $S_2 \equiv - \log_2 \left({\rm tr}\left( \rho_A{(\theta,\phi)} ^2 \right)\right)$.  A detailed derivation is provided in \ref{appendix}. In the end, we find that
\beqa
{\rm tr}\left( \rho_A{(\theta,\phi)} ^2 \right) &=& |a|^4 + \frac{|b|^4 d_A}{f} + \frac{(a^* b)^2}{|G|} + \frac{(a b^*)^2}{|G|} \nonumber \\
&+&  \frac{2|a|^2 |b|^2}{f}  + \frac{2 |a|^2 a^*b}{\sqrt{|G|}} + \frac{2 |a|^2 a b^*}{\sqrt{|G|}}  \nonumber \\
&+&   \frac{2 a^* |b|^2 b ~ d_A}{f \sqrt{|G|}} +  \frac{2a |b|^2 b^*  d_A}{f \sqrt{|G|}} + \frac{2 |a|^2 |b|^2 d_A}{\sqrt{|G|}}, 
\label{thisone} 
\eeqa
from which $S_2$ follows. The above expression holds for any size of subsets $A$ and $B$ for a state of the type in equation~(\ref{esta}). Using this equation, one can verify that all states except the state  $\ket{\bar{0}}$ are entangled (using the entropy of the reduced state); as a result, our Bloch sphere must be in ``Class 4'' according to the classification in \cite{bloch}: only one state is separable, and all the other states, including those mixed states in the interior, are entangled.
 
\section{Topological order on the Bloch sphere}
Our approach to assess topological order is to study the subleading correction to the 2-R\'enyi entropy of a contiguous block of spins. For non-chiral two-dimensional topological order, all such possible corrections have been classified and correspond to a \emph{discrete} set of values \cite{levinwen}.

Thus, we now particularize the scenario to the case of a contiguous block of spins. Following the notation in \cite{hamma}, we denote $n_s = \Sigma_A + \Sigma_B + \Sigma_{AB}$, where $\Sigma_{A,B}$ is the number of star operators acting \emph{only} on $A,B$ (respectively), and $\Sigma_{AB}$ is the number of those acting on \emph{both} $A$ and $B$. For a block of spins, $\Sigma_{AB}$ is clearly proportional to the boundary between the block ($A$) and the rest of the system ($B$). It is also easy to see that $d_{A,B} = 2^{\Sigma_{A,B}}$, $|{G}| = 2^{n_s-1}$, and $f^{-1} = d_B/|G| = 2^{-(n_s - 1) + \Sigma_B} = 2^{- \Sigma_{AB} + 1 - \Sigma_A}$.  Thus, fixing $\Sigma_{A,B}$ and $\Sigma_{AB}$ fully determines the total number of spins both in the system and in the block. For the sake of simplicity, from now on we focus on the case of a lattice on a torus with $k \times k$ sites and a block (A) with $L \times L$ sites; however, we remark that other more general cases can be analyzed similarly, leading to similar conclusions. One then gets
\beq
\Sigma_A = L^2,  ~~ \Sigma_B = k^2 - L^2 - 4L, ~~\Sigma_{AB} = 4L,  
\eeq
where we took the convention that the spins sitting at the links of the boundary of a block of sites are considered as being  \emph{inside} the block. 

Combining all the above expressions, we can now compute the 2-R\'enyi entropy of an arbitrary block of spins. In particular, we find that the trace of the squared reduced density matrix is
\beqa
{\rm tr}\left( \rho_A{(\theta,\phi)} ^2 \right) &=& \left(\sin{\left(\frac{\theta}{2}\right)}\right)^4  \frac{\left(2^{2 k^2 - 4L - 1} - \left(2^{L^2 + 1} - 1\right)  \left(2^{k^2 - L^2 - 4L + 1} - 1\right)\right)}{\left(2^{k^2 - 1} - 1\right)^2} \nonumber \\
&+& 4 \cos{(\phi)} \left(\sin{\left(\frac{\theta}{2}\right)}\right)^3 \cos{\left(\frac{\theta}{2}\right)} \frac{\left(2^{L^2} - 1\right) \left(2^{k^2 - L^2 - 4L} - 1\right)}{\left(2^{k^2 - 1} - 1\right)^{3/2}} \nonumber \\
&+& 2 \left(\sin{\left(\frac{\theta}{2}\right)}\right)^2 \left(\cos{\left(\frac{\theta}{2}\right)}\right)^2 \frac{\left(2^{L^2} + 2^{k^2 - L^2 - 4L} - 2\right)}{\left(2^{k^2 - 1} - 1\right)} \nonumber \\
&+& \left(\cos{\left(\frac{\theta}{2}\right)}\right)^4. 
\label{longkl}
\eeqa

Let us study equation~(\ref{longkl}) in some detail in order to find whether topological order exists or not for different values of $\theta$ and $\phi$. First, we take the limit of a large (possibly infinite) lattice, $k^2 \gg 1$. In this limit we find
\beqa
{\rm tr}\left( \rho_A{(\theta,\phi)} ^2 \right) &\approx&  \left(\sin{\left(\frac{\theta}{2}\right)}\right)^4 2^{-4L+1}
+ 2 \left(\sin{\left(\frac{\theta}{2}\right)}\right)^2 \left(\cos{\left(\frac{\theta}{2}\right)}\right)^2 2^{-L^2-4L+1} \nonumber \\
&+& \left(\cos{\left(\frac{\theta}{2}\right)}\right)^4. 
\eeqa
The above equation is independent of $\phi$, which means that the entropy is almost independent of this angle for large lattices (and, in fact, completely independent for the infinite-size lattice). Next, we take the limit of a large block, i,.e., $L^2 \gg L \gg 1$ (but still $k^2 \gg L^2$). In this limit we find 
\beq
{\rm tr}\left( \rho_A{(\theta,\phi)} ^2 \right) \approx  \left(\sin{\left(\frac{\theta}{2}\right)}\right)^4 2^{-4L+1} +  \left(\cos{\left(\frac{\theta}{2}\right)}\right)^4. 
\label{kldos}
\eeq
We can now evaluate this expression for the toric code's ground state. This state is given by $\theta = \theta_0 = 2 \arccos{(1/\sqrt{|{G}|})}$ with $|G| = 2^{k^2 - 1}$. Since we are in the limit of an infinite lattice ($k^2 \gg 1$), this means that 
\beq
\left(\sin{\left(\frac{\theta}{2}\right)}\right)^4 \approx 1, ~~~ \left(\cos{\left(\frac{\theta}{2}\right)}\right)^4 \approx 0,  
\eeq
and therefore 
\beq
S_2 \approx 4L - 1, 
\eeq
which is the known result for the 2-R\'enyi entropy of the toric code \cite{hamma}. As a common rule, the topological contribution $S_\gamma$ is defined by the scaling $S_2(L) = \alpha L + S_\gamma + O(1/L)$, where $\alpha$ is some area-law prefactor. Importantly, in this paper we take this scaling law as our ``practical definition'' of a topologically-ordered state. As proven in \cite{toporen}, this is the correct scaling behavior with $L$ for the 2-R\'enyi entropy at a topological phase in two spatial dimensions \footnote{See also \cite{topovn, topogeo} for similar scalings of other entanglement measures as well as \cite{pachos} for a detailed discussion on topologically-ordered states.}. Thus, we obtain the known result $S_\gamma = -1$ for the toric code's ground state $\ket{\Psi_0}$.

\begin{figure}
	\centering
	\includegraphics[width=\linewidth]{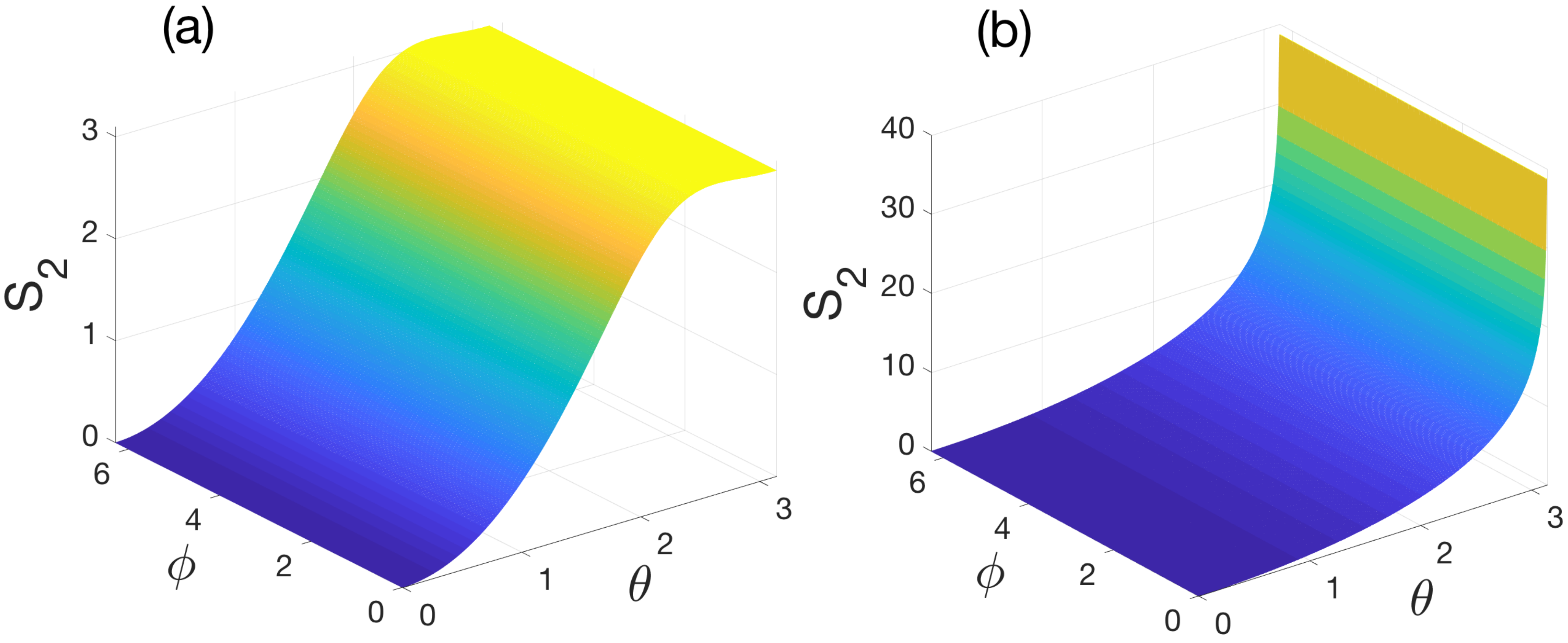}
	\caption{2-R\'enyi entropy in the $\theta-\phi$ plane: (a) $k=20, L=1$; (b) $k=20, L=10$. The dependence with $\phi$ is extremely weak.}
	\label{fig2}
\end{figure}

We now prove that small perturbations in the angle $\theta$ around the toric code's ground state destroy topological order. To see this, consider 
{
\beq
\left(\sin{\left(\frac{\theta}{2}\right)}\right)^4 \approx 1-2\sqrt{\epsilon} + \epsilon, ~~~ \left(\cos{\left(\frac{\theta}{2}\right)}\right)^4 \approx \epsilon,  
\label{eqapp}
\eeq}
with $\epsilon \ll 1$ \footnote{{If $(\cos{\alpha})^4 \approx \epsilon$  and $(\cos{\alpha})^2 + (\sin{\alpha})^2 = 1$, then $(\sin{\alpha})^2 \approx 1 - \sqrt{\epsilon}$, and therefore $(\sin{\alpha})^4 \approx 1 - 2\sqrt{\epsilon} + \epsilon$.}}. Using equation~(\ref{kldos}) and assuming $\epsilon   \ll 2^{-4L}$ (indeed \emph{very} small if $L \gg 1$), one finds (up to $O(\epsilon)$ terms)
\beq
S_2 \approx 4L - 1 + \frac{2 \sqrt{\epsilon}}{\ln 2},  
\label{subl}
\eeq
which means that any possible constant topological correction would depend on $\epsilon$,  thus not being in the toric code universality class \footnote{To get this expression, we kept $O(\sqrt{\epsilon})$ terms, and used also $\ln (1- 2 \sqrt{\epsilon}) \approx - 2 \sqrt{\epsilon}$. For finite lattices, a similar result is found for perturbations in $\phi$.}.

In order to further verify our results, we computed numerically the 2-R\'enyi entropy of an arbitrary block of spins, directly using equation~(\ref{longkl}) and before taking any approximation: see figures \ref{fig2} and \ref{fig3} for some examples. The plots are consistent with our analytical results. First, there is \emph{only one} separable state in the whole Bloch sphere (the state $\ket{\bar{0}}$, for which the azimuthal angle $\phi$ plays no role). Second, for fixed $k$ and $L$, we notice that the R\'enyi entropy attains a maximum as a function of $\theta$, as illustrated in figure \ref{fig3}. This maximum is \emph{not} related to the angle $\theta_0$ at which we have the toric code's ground state; in fact, it drifts towards the right as $L$ increases. Third, we can extract the topological contribution $S_\gamma$ to the 2-R\'enyi entropy from the scaling $S_2(L) = \alpha L + S_\gamma + O(1/L)$, where $\alpha$ is some area-law prefactor. From figure \ref{fig4} we can see that $S_\gamma = -1$ \emph{only} for the point corresponding to the toric code ground state $\ket{\Psi_0}$, and that it changes continuously with angles $\theta$ and $\phi$. These numerical checks reinforce our result that, on the surface of this Bloch sphere, there seems to be only one topological state, according to the conventional defining properties of a topological phase.

\begin{figure}
	\centering
	\includegraphics[width=\linewidth]{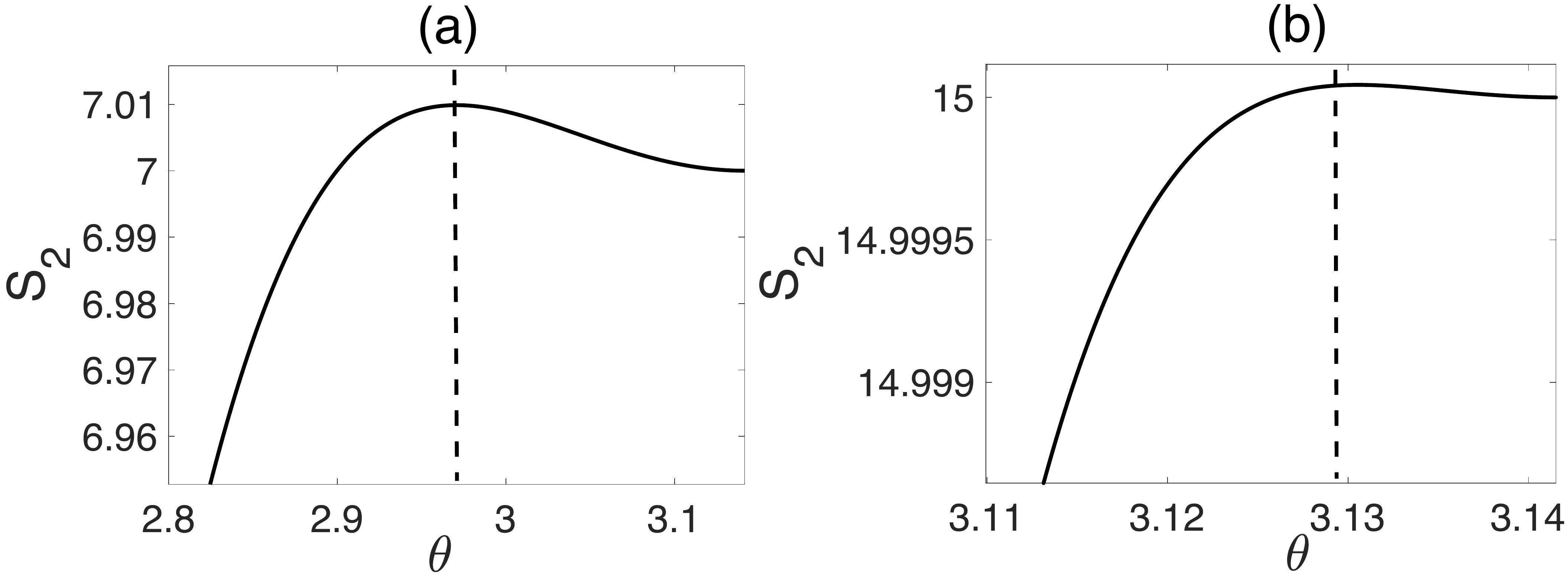}
	\caption{ 2-R\'enyi entropy for $\phi = 0$ as a function of $\theta$: (a) $k = 20, L = 2$; (b) $k = 20, L = 4$. The position of the maximum (dotted line) shifts as a function of $L$.}
	\label{fig3}
\end{figure}
\begin{figure}
	\centering
	\includegraphics[width=\linewidth]{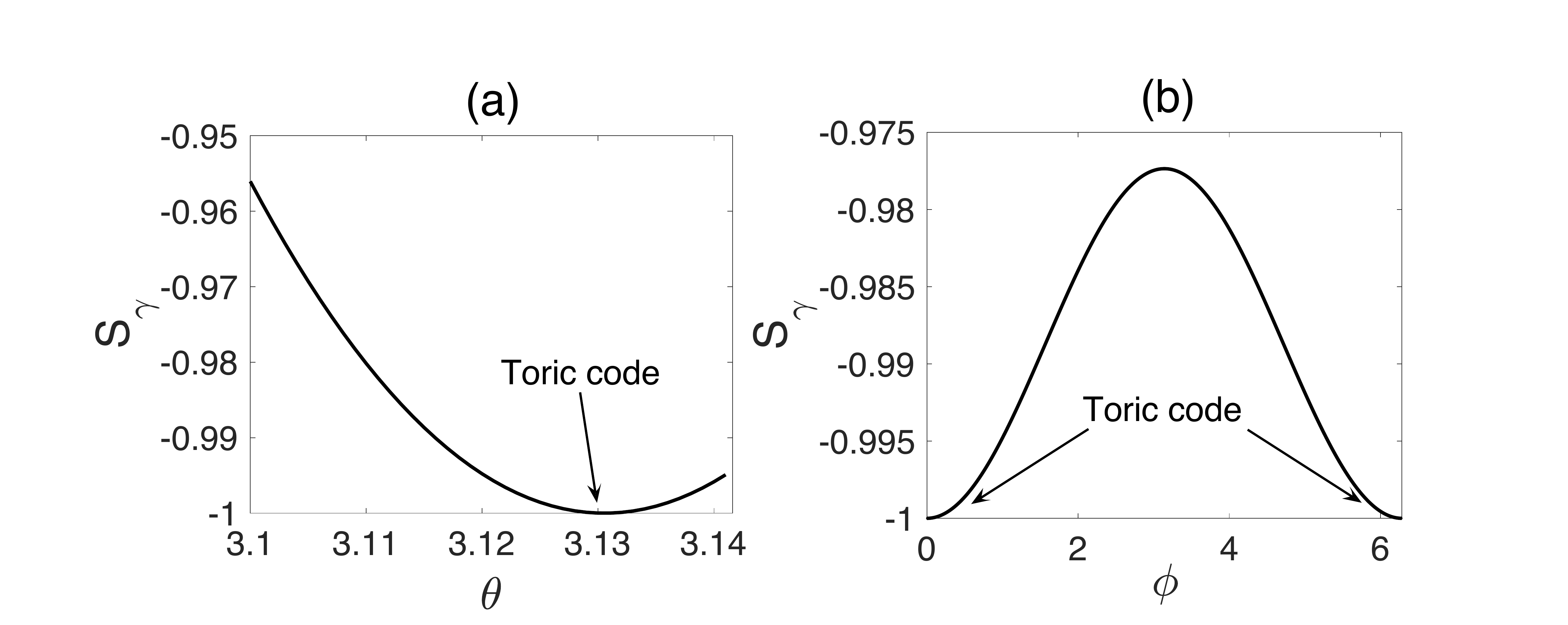}
	\caption{Topological contribution to the 2-R\'enyi entropy of a block, for a system with $k=4$: (a) at $\phi = \phi_0 = 0$ and as a function of $\theta$; (b) at $\theta = \theta_0 = 2 \arccos{\left( 1 / \sqrt{|G|} \right)}$ and as a function of $\phi$. The topological contribution of the toric code universality class $S_\gamma = -1$ is only recovered for the toric code point. The dependence with the angles gets weaker as $k$ increases.}
	\label{fig4}
\end{figure}

Formally, it is not impossible that other topological universality classes may be reached continuously. However, there are several counterarguments for this possibility. For instance, the subleading correction in equation~(\ref{subl}) seems to have a complicated dependence, which is unusual for topological states. Moreover, the resulting state for any finite $\epsilon$ does not seem to have (at least a priori) a \emph{gauge symmetry}, which is also known to be a defining property of topological order \cite{chen}. We conclude, thus, that \emph{formally} there seems to be only one topological state on the surface of this Bloch sphere for $O(\epsilon)$ perturbations, which is the toric code ground state \footnote{We remind the reader that $S_\gamma = -1$ corresponds to the topological universality class of the toric code model, while close values (even $S_\gamma = -0.9999$) do not seem to correspond to any non-chiral two-dimensional topological universality class \cite{levinwen}. In other words, a state with $S_\gamma = -0.9999$ would very probably flow to a trivial product state under coarse-grainings, which is a defining property of a non-topological state.}. Still, we remark that our $O(\epsilon)$ analysis cannot say anything for tiny neighborhoods around the toric code point, since for analyzing them we should keep all orders in $\epsilon$, which we did not do. A numerical analysis would also be of no help here, due to accuracy problems when analyzing tiny infinitesimal environments. The situation in the interior of the Bloch sphere will be briefly discussed in the conclusions.

\section{Topological order and Grover rotations}
We now come back to the expressions in equations (\ref{gr1}) and (\ref{gr2}). Indeed, these expressions have exactly the form of a Grover rotation. To be more specific, our setting is equivalent to that of a quantum search algorithm trying to find the state $\ket{\bar{0}}$ within an unstructured database of the $|G|$ elements $\{g\ket{0}^{\otimes n}\}_{g \in G}$: this is actually a quantum search within a group $G$. In our case, the ground state $\ket{\Psi_0}$ is the equally-weighted superposition of all the elements in the database. The so-called \emph{Grover kernel} is defined as
\beq
K \equiv (2\ket{\Psi_0} \bra{\Psi_0} - \mathbb{I})O,\label{eq_grov_ker}
\eeq
where $O$ is the so-called \emph{quantum oracle}. It is proved in \ref{appendix} that by applying $K$ to the state $\ket{\Psi_0}$ a number of times $O(\sqrt{|G|})$, one can rotate $\ket{\Psi_0}$ approximately into the desired searched state $\ket{\bar{0}}$. Similarly, by applying the inverse Grover kernel $K^{-1}$ to the product state $\ket{\bar{0}}$ a number of times $O(\sqrt{|G|})$, one actually rotates this product state $\ket{\bar{0}}$ approximately into the topologically-ordered state $\ket{\Psi_0}$. Full details and conclusions are available in \ref{appendix}.

\section{Conclusions and further remarks}
In this paper we have shown that two-dimensional non-chiral topological order does not seem to survive some Bloch-sphere perturbations. This was shown by constructing a Bloch sphere from one of the ground states of Kitaev's toric code model together with one of its closest product states. We studied the entanglement of the pure states on the surface of this sphere and concluded that there is only one product state and that there seems to be, at least formally, only one topologically-ordered state. We have also shown that topologically-ordered states may be approximated as states arising from a quantum search with a gauge symmetry, and that such an approximation may be very good for large systems.

The results in this paper can be generalized in different ways. For instance, it is possible that similar results could be obtained for more general topological models with other gauge groups~\cite{wen}; however, proving (or disproving) such generalizations is not obvious, and we leave it as a topic for future research. It would also be interesting to study other types of Bloch spheres built from topological states. In this sense, the sphere obtained from two orthogonal toric code ground states is indeed the class of states analyzed in \cite{oshik}, which motivated the definition of Minimally Entangled States (MES). It is easy to see that such a Bloch sphere does not contain any product state, and therefore falls into ``Class 5'' according to \cite{bloch}. In fact, as proven in \cite{oshik}, all pure states in such a Bloch sphere are topologically-ordered, since all of them are valid ground states of the toric code's Hamiltonian.

\begin{figure}
	\centering
	\includegraphics[width=0.61\linewidth]{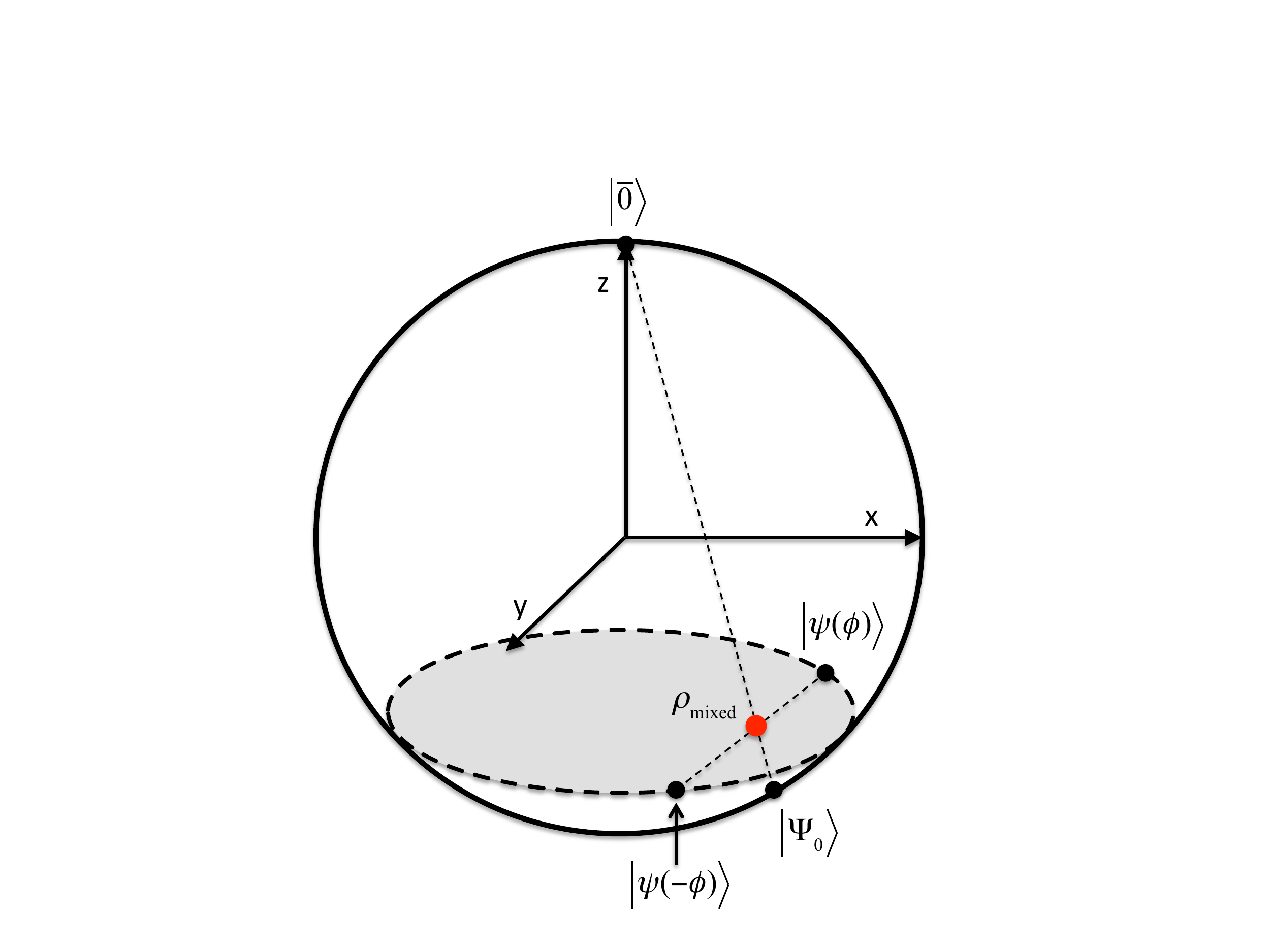}
	\caption{A specific rank-2 mixed state $\rho_{\mathrm{mixed}}$ presented in its relevant Bloch sphere: this state is very close to the topological state $\ket{\Psi_0}$, and it is a mixture of the topological state $\ket{\Psi_0}$ and the tensor product state $\ket{\overline{0}}$ (giving a very high probability to $\ket{\Psi_0}$). The two pure states $\ket{\psi(\phi)}$ and $\ket{\psi(-\phi)}$ are chosen such that both are placed in the same $x$ and $z$ coordinates as the mixed state $\rho_{\mathrm{mixed}}$ (although they differ in the $y$ coordinate), and their 50\%-50\% mixture is another description of the same mixed state $\rho_{\mathrm{mixed}}$.}
	\label{fig5}
\end{figure}

Finally, one can ask: what about the interior of the sphere? This question is a bit controversial. Imagine that one has a classical mixture, with $99.99 \%$ of a topological state and $0.01 \%$ of a perturbation that formally breaks topological order (see figure \ref{fig5}). Such state may not be considered as topological (no gauge symmetry, incorrect topological entropy, flow towards a product state under coarse-grainings, etc.); however, measurements over this state would produce results compatible with topological order $99.99 \%$ of the time.
In fact, similar controversial questions may be asked about a mixture of
two (or more) ``almost-topological'' pure states (e.g., the two pure states presented in figure \ref{fig5}), where an
``almost-topological'' pure state is a non-topological state that is very
close to a topological state. Those questions can be relevant because if we
carefully choose two such ``almost-topological'' pure states, their mixture
can be exactly the same mixed state as described above (namely, a mixture of
a topological state with a small non-topological perturbation).
Are such states topological in practice, or not? We believe that clarifying this is quite important because, in actual real-life experiments, one will \emph{most probably never} reproduce a topological state with $100 \%$ accuracy, but rather a mixture of such state with (perhaps non-local) perturbations. In our opinion, this is an indication that one may need a more accurate, practical and relaxed definition of a topological state in such settings.

We may also take a more computer-science-oriented approach, asking whether a trivial circuit \cite{Hastings11,Hastings13,AharonovTouati18} can be used for yielding a specific state, when starting with a trivial (namely, tensor product) state; note that a trivial circuit is a quantum circuit with constant depth and range. A plausible conjecture is that the mixed state $\rho_{\mathrm{mixed}}$ (and potentially also the pure states $\ket{\psi(\phi)}$ and $\ket{\psi(-\phi)}$) in figure \ref{fig5} cannot be obtained using a trivial circuit, neither from the tensor product state $\ket{\overline{0}}$ nor from the topological state $\ket{\Psi_0}$.

If this conjecture is correct, then there would be (at least) three different regions on the Bloch sphere: a single non-topological product state, a single topological state, and a continuum of non-topological states that cannot be obtained from any product state using a trivial circuit (namely, they are ``non-trivial'' states in the sense defined by~\cite{Hastings13,BHV06}). In addition, there would be at least two phase transitions: a phase transition that singles out the tensor product state $\ket{\overline{0}}$ from all its neighboring states on this Bloch sphere, and another phase transition that singles out the topological state $\ket{\Psi_0}$ from all its neighboring states. Therefore, the Bloch sphere seems to consist almost only of states that are neither ``trivial'' states, nor topologically ordered.

The above conjecture seems plausible, because the Grover kernel $K$ (defined in equation~(\ref{eq_grov_ker})), which corresponds to a rotation on the $X$-$Z$ plane of the Bloch sphere, seems highly non-trivial to implement (in particular, it seems very difficult to implement the projector $\ket{\Psi_0} \bra{\Psi_0}$ using a trivial circuit). Moreover, since \emph{all} rotations on the $X$-$Z$ plane of the Bloch sphere are powers of the Grover kernel $K$ (as noted in \ref{appendix}) and seem hard to implement, we can further conjecture that any two points on the $X$-$Z$ plane cannot be rotated to one another using a trivial circuit, which means that there is an \emph{infinite} number of regions on the Bloch sphere that are \emph{unreachable} from each other using trivial circuits. We leave a deep analysis of those conjectures and their conclusions as open problems for future works. We also leave the analysis of mixed states defined inside the Bloch sphere for future work.

\ack
The authors thank Itai Arad for useful discussions. R. O. also acknowledges Itai Arad for his invitation and hospitality in Technion. The work of T.M. and R.L. was partly supported by the Israeli MOD Research and Technology Unit.

\appendix

\section{\label{appendix}}

\subsection{Details on the analytical derivation of the 2-R\'enyi entropy}

In this appendix we provide details about the derivation of the 2-R\'enyi entropy. In equation~(\ref{thisone}) of the main paper, the term $ {\rm tr} \left( \rho_A(\theta, \phi)^2 \right)$ is particularly important since  from it we can obtain the exact 2-R\'enyi entropy of the system for any size of subsets $A$ and $B$. Let us now see more precisely how this is derived. 

First, one computes $\rho_A(\theta, \phi)$, the reduced density matrix of a subset. This is defined as $\rho_A{(\theta,\phi)} \equiv {\rm tr}_B \left( \ket{\theta, \phi} \bra{\theta, \phi} \right)$, i.e., the partial trace over $B$ on the projector for state $\ket{\theta, \phi}$. We notice that we can write this state as 
\beq
\ket{\theta, \phi} = a \ket{\bar{0}} + b \ket{\Psi_0}, 
\eeq
where $a$ and $b$ are given by equation~(\ref{ab}) in the main text, $\ket{\Psi_0}$ is the toric code ground state in equation~(\ref{tcstate}), and $\ket{\bar{0}} \equiv \ket{0}^{\otimes n}$ ($n$ is the number of spins). Using this, the projector then reads 
\beq
\ket{\theta, \phi} \bra{\theta, \phi} = |a|^2 \ket{\bar{0}} \bra{\bar{0}} + |b|^2 \ket{\Psi_0}\bra{\Psi_0} + a^*b \ket{\Psi_0}\bra{\bar{0}} + b^*a \ket{\bar{0}}\bra{\Psi_0}. 
\eeq
We now take the partial trace over $B$. For the first term in the above equation, this gives ${\rm tr}_B \left( \ket{\bar{0}} \bra{\bar{0}} \right) = \ket{0_A} \bra{0_A}$, where $\ket{0_A} \equiv \ket{0}^{\otimes n_A}$, and $n_A$ is the number of spins in $A$. For the second term, we obtain exactly the reduced density matrix for the toric code ground state. This has been computed explicitly in theorem 1 of \cite{hamma}, and gives
\beq
{\rm tr}_B \left( \ket{\Psi_0} \bra{\Psi_0} \right) = \rho_A^{(0)} = \frac{d_A}{f} P_A, 
\eeq
where $d_A$ and $f$ are defined as in the main text, and $P_A$ is a projector on a subspace of dimension $f d_A^{-1}$. For the third and fourth terms in the partial trace, the calculation is trickier. The key point is to notice that 
\beq
{\rm tr}_B \left(  \ket{\bar{0}}\bra{\Psi_0} \right) = \braket{0_B}{\bar{0}} \braket{\Psi_0}{0_B} = \frac{1}{\sqrt{|G|}} \ket{0_A} \bra{\phi_A}, 
\eeq
where  
\beq 
\ket{\phi_A} \equiv \sum_{g \in G_A} g_A \ket{0_A} 
\eeq
is a non-normalized quantum state, and $\ket{0_B} \equiv \ket{0}^{n_B}$ ($n_B$ being the number of spins in $B$). Thus, at the end of the day, the reduced density matrix for subsystem $B$ reads 
\beqa
\rho_A{(\theta,\phi)} &\equiv& {\rm tr}_B \left( \ket{\theta, \phi} \bra{\theta, \phi} \right) = |a|^2 \ket{0_A}\bra{0_A} + |b|^2 \rho_A^{(0)} \nonumber \\
&+& \frac{1}{\sqrt{|{G}|}}\left(a^*b \ket{\phi_A} \bra{0_A} +  b^*a \ket{0_A} \bra{\phi_A}\right), 
\eeqa
which is the expression given in the main text.

Having this reduced density matrix, one can compute all bipartite entanglement properties of the state $\ket{\theta, \phi}$. In this paper we focus on the 2-R\'enyi entropy, since it is simpler to calculate than other measures of entanglement. For this, the next step is to take the square of $\rho_A{(\theta,\phi)}$, and then the trace. The square, though tedious, is not difficult to compute. However, the delicate point in this calculation is the trace. In order to compute it, the key is to use the following relations: 
\beqa
\left( \rho_A^{(0)} \right)^2 = \frac{d_A}{f}  \rho_A^{(0)}, &&{\rm tr}\left(  \rho_A^{(0)} \right) = 1, \\
\braket{0_A}{0_A} = 1,&  &\braket{\phi_A}{\phi_A} = |G_A|, \\
\braket{\phi_A}{0_A} = 1, && \braket{0_A}{\phi_A} = 1, \\
\bra{0_A} \rho_A^{(0)} \ket{0_A} = \frac{1}{f} \braket{0_A}{\phi_A} = \frac{1}{f}, && \bra{0_A} \rho_A^{(0)} \ket{\phi_A} = \frac{1}{f} \braket{\phi_A}{\phi_A} = \frac{|G_A|}{f}. 
\eeqa
The last line above follows from the property $\rho_A^{(0)} \ket{0_A} = f^{-1} \ket{\phi_A}$, which is proven as follows: 
\beqa
\rho_A^{(0)} \ket{0_A} &=& \frac{1}{f} \sum_{g \in G/G_B \atop \tilde{g} \in G_A} g_A \ket{0_A}\bra{0_A} g_A \tilde{g}_A \ket{0_A} = \frac{1}{f} \sum_{g \in G/G_B \atop \tilde{g} \in G_A} g_A \ket{0_A} \delta_{g_A, \tilde{g}_A} \nonumber \\
&=& \frac{1}{f} \sum_{g \in G_A} g_A \ket{0_A} = \frac{1}{f} \ket{\phi_A}, 
\eeqa
where we used the expression 
\beq
\rho_A^{(0)}=  \frac{1}{f} \sum_{g \in G/G_B \atop \tilde{g} \in G_A} g_A \ket{0_A}\bra{0_A} g_A \tilde{g}_A
\eeq
for the toric code's reduced density matrix, as explicitly derived in \cite{hamma}. Using all these relations one can arrive finally (and after simplification) to equation~(\ref{thisone}) in the main text for the trace of the square of $\rho_A{(\theta,\phi)}$, and from there to the 2-R\'enyi entropy.

\subsection{Topological order and Grover rotations} 

We now come back to the expressions in equations (\ref{gr1}) and (\ref{gr2}) in the main text. Indeed, these expressions have exactly the form of a Grover rotation. To be more specific, our setting is equivalent to that of a quantum search algorithm trying to find the state $\ket{\bar{0}}$ within an unstructured database of the $|G|$ elements $\{g\ket{0}^{\otimes n}\}_{g \in G}$: this is actually a quantum search within a group $G$. In our case, the ground state $\ket{\Psi_0}$ is the equally-weighted superposition of all the elements in the database. The so-called \emph{Grover kernel} is defined as 
\beq
K \equiv (2\ket{\Psi_0} \bra{\Psi_0} - \mathbb{I})O, 
\eeq
where $O$ is the so-called \emph{quantum oracle}. This oracle is defined such that $O\ket{x} = (-1)^{f(x)}\ket{x}$, where $f(x) = 0$ if $x \neq \bar{0}$ and $f(x) = 1$ if $x = \bar{0}$; thus, it shifts the phase of the searched element $\ket{\bar{0}}$ only \cite{grover}. A well-known result in quantum computation is that successive applications of the kernel $K$ on the state $\ket{\Psi_0}$ are equivalent (in the limit of a large $|G|$) to the rotation 
\beq
K^m \ket{\Psi_0} = \sin{\left( \frac{2m+1}{2} \tilde{\theta} \right)} \ket{\bar{0}} + \cos{\left( \frac{2m+1}{2} \tilde{\theta} \right)} \ket{\bar{1}}, 
\eeq
where $\tilde{\theta} \equiv 2 \arcsin{\left( 1/\sqrt{|G|} \right)} \approx 2/\sqrt{|G|}$. Thus, for a number of iterations $m = O(\sqrt{|G|})$, one can rotate $\ket{\Psi_0}$ approximately into the desired searched state $\ket{\bar{0}}$.

Notice that one can also rotate backwards in a similar way: by applying the inverse Grover kernel $K^{-1}$ to the product state $\ket{\bar{0}}$ a number of times $O(\sqrt{|G|})$, one actually rotates this product state $\ket{\bar{0}}$ approximately into the topologically-ordered state $\ket{\Psi_0}$. This is important because it offers a way to prepare \emph{approximate} topologically-ordered states on quantum computers capable of running Grover's quantum search algorithm with gauge symmetry (the gauge symmetry is $\mathbb{Z}_2$ for the toric code). Notice, however, that the \emph{exact} topological state cannot usually be prepared using this procedure, because Grover's iterations involve \emph{discrete} jumps in the rotation angle; therefore, in general, one cannot fine-tune the topological state exactly. However, for large systems, one can get extremely (exponentially!) close to the topological state, which may also be useful in practice.

We further note that \emph{all} rotations on the $X$-$Z$ plane of the Bloch sphere (which includes $\ket{\bar{0}}$, $\ket{\bar{1}}$, and $\ket{\Psi_0}$) are \emph{real powers} of the Grover kernel $K \equiv (2\ket{\Psi_0} \bra{\Psi_0} - \mathbb{I})O$. In other words, the real powers $\{K^\epsilon \mid \epsilon \in \mathbb{R}\}$ characterize all the possible rotations from one state to the other on the $X$-$Z$ plane of the Bloch sphere.

As a remark, notice that the states in Grover's algorithm for $n$ qubits can also be described in the Bloch sphere formalism, although in that case the Bloch sphere falls into ``Class 3'' according to the classification in \cite{bloch}: namely, there are two pure product states (not orthogonal to each other) on the Bloch sphere. This is true because in the usual Grover's setting, if no other restrictions are applied, there are two product states: the superposition of all the database elements for $n$ qubits and the searched state. However, in our case, the superposition of all database elements is not a product state, and it is even a topologically-ordered entangled state.


\section*{References}
\begin{thebibliography}{}
\bibitem{wen}
X.-G- Wen, \emph{Quantum Field Theory of Many-Body Systems}, Oxford University Press, USA, 2004. 

\bibitem{lgt}
J. Kogut, L. Susskind, Phys. Rev. D {\bf 11}, 395 (1975). 

\bibitem{kita}
A. Y. Kitaev, Annals of Physics {\bf 303}, 2-30 (2003).

\bibitem{robustTO}
A. Hamma D. A. Lidar, Phys. Rev. Lett. {\bf 100}, 030502 (2008); S. Trebst \emph{et al.}, Phys. Rev. Lett. {\bf 98}, 070602 (2007);  I. S. Tupitsyn \emph{et al.}, Phys. Rev. B {\bf 82}, 085114 (2010);  J. Vidal, S. Dusuel, K. P. Schmidt, Phys. Rev. B {\bf 79}, 033109 (2009);  J. Vidal \emph{et al.}, Phys. Rev. B {\bf 80}, 081104(R) (2009); J. Yu, S.-P. Kou, X.-G. Wen, Europhys. Lett. {\bf 84}, 17004 (2008); X.-G. Wen, Phys. Rev. Lett. {\bf 90}, 016803 (2003); S. Dusuel \emph{et al.}, Phys. Rev. Lett. {\bf 106}, 107203 (2011);  L. Tagliacozzo, G. Vidal, Phys. Rev. B {\bf 83}, 115127 (2011); M. D. Schulz \emph{et al.}, New J. Phys. {\bf 14}  025005 (2012); H. He, H. Moradi, X.-G. Wen, Phys. Rev. B {\bf 90}, 205114 (2014); S. Bravyi, M. Hastings, S. Michalakis, J. Math. Phys. {\bf 51}, 093512 (2010); S. Michalakis, M. Zwolak,  Commun. Math. Phys. {\bf 322}, pp. 277-302 (2013).

\bibitem{tempTO}
S. Iblisdir \emph{et al.}, Phys. Rev. B {\bf 79}, 134303 (2009). 

\bibitem{TOMBL}
B. Bauer, C. Nayak, J. Stat. Mech. P09005 (2013). 

\bibitem{closest}
R. Or\'us \emph{et al.}, New J. Phys. {\bf 16}, 013015 (2014);  R. Or\'us \emph{et al.}, Phys. Rev. Lett. {\bf 113}, 257202 (2014); O. Buerschaper \emph{et al.}, JSTAT P11009 (2014). 

\bibitem{bloch}
M. Boyer, R. Liss, T. Mor, Phys. Rev. A {\bf 95}, 032308 (2017). 

\bibitem{BHV06}
S. Bravyi, M.~B. Hastings, and F. Verstraete, Phys. Rev. Lett. \textbf{97}, 050401 (2006).

\bibitem{Hastings13}
M.~B. Hastings, Quantum Info. Comput. \textbf{13}, 393 (2013).

\bibitem{grover}
L. K. Grover, Proceedings, 28th Annual ACM Symposium on the Theory of Computing, (May 1996) p. 212; M. Boyer \emph{et al.}, Fortsch. Phys. {\bf 46}: 493-506,1998.

\bibitem{hamma}
A. Hamma, R. Ionicioiu, P. Zanardi,  	Phys. Rev. A {\bf 71}, 022315 (2005). 

\bibitem{levinwen}
M. A. Levin, X.-G. Wen, Phys. Rev. B {\bf 71}, 045110 (2005). 

\bibitem{toporen}
S. T. Flammia \emph{et al.}, Phys. Rev. Lett. {103}, 261601 (2009)

\bibitem{topovn}
A. Kitaev, J. Preskill, Phys. Rev. Lett. {\bf 96}, 110404 (2006); M. Levin, X.-G. Wen, Phys. Rev. Lett. {\bf 96}, 110405 (2006). 

\bibitem{topogeo}
R. Or\'us \emph{et al.}, New J. Phys. {\bf 16}, 013015 (2014); R. Or\'us \emph{et al.}, Phys. Rev. Lett. {\bf 113}, 257202 (2014); O. Buerschaper \emph{et al.}, JSTAT P11009 (2014). 

\bibitem{pachos}
J. K. Pachos, Introduction to Topological Quantum Computation, Cambridge University Press (2012). 

\bibitem{chen} 
X. Chen \emph{et al.}. Phys. Rev. B {\bf 82}, 165119 (2010).

\bibitem{oshik}
Y. Zhang \emph{et al.}, Phys. Rev. B {\bf 85}, 235151 (2012). 

\bibitem{Hastings11}
M.~B. Hastings, Phys. Rev. Lett.  \textbf{107}, 210501 (2011).

\bibitem{AharonovTouati18}
D. Aharonov and Y. Touati, arXiv preprint arXiv:1810.03912 (2018).

\end{thebibliography}
\end{document}